\documentstyle[12pt]{article}
\begin{document}
\begin{center}
{\Large \bf THE EQUIVALENCE THEOREM FOR CHIRAL LAGRANGIANS}  \\
\vskip 3.0cm
{\large \ Antonio DOBADO \footnote{E-mail: dobado@cernvm.cern.ch}
and Jos\'e Ram\'on PELAEZ \footnote {E-mail:pelaez@fis.ucm.es }   \\
\vskip 1.0cm
 Departamento de F\'{\i}sica Te\'orica  \\
 Universidad Complutense de Madrid\\
 28040 Madrid, Spain \\
\vskip 0.5cm
\large FT/UCM/2/94 \\
\large hep-ph 9404239 }
 \vskip 1.0cm
{\bf \large April 1994}
\vskip 1.0cm
\begin{abstract}
In this
work we derive the version of the Equivalence Theorem that applies when the
symmetry breaking sector of the Standard Model is described by a general
chiral lagrangian. The demonstration is valid  for renormalized fields for any
value of the gauge parameter (in $R_{\xi}$ gauges)  and any parametrization of
the coset space. It is based in the absence of gauge anomalies which makes it
possible to build  an (anti)-BRS invariant chiral lagrangian in terms of the
renormalized fields  and therefore to use the corresponding Ward identities to
obtain the theorem. \end{abstract} \end{center}
 \newpage

\section{Introduction}

This paper deals with the problem of finding a general derivation of the
so-called Equivalence Theorem (ET) [1,2,3] relating the $S$ matrix elements of
electroweak gauge bosons longitudinal  components  with the corresponding $S$
matrix  elements for the would be Goldstone Bosons (GB) at high energies
compared with the  electroweak symmetry breaking scale $v=250 GeV$. This kind
of relation could be very useful  in order to obtain information about the
Symmetry Breaking Sector (SBS) of the  Standard Model (SM) from the future LHC
data since computations are by far easier to be done for scalars than for
longitudinal gauge bosons. Despite the very precise data  collected at LEP we
have virtually no information about the SBS of the SM and it is still a
mystery whether this sector can be described by the Minimal Standard Model
(MSM) with just one doublet of Higgs fields, the Minimal Supersymmetric SM
(see [4] for a review), Technicolor [5] or other models like the so-called BESS
model [6].

As it is not possible, at present, to know which is the dynamics responsible
for the spontaneous breaking of the electroweak group $SU(2)_L\times U(1)_Y$
to  the electromagnetic group $U(1)_{em}$, it is  very important to develop
some model independent framework to describe, at least phenomenologically, the
SBS mechanism. Recently such  a framework has been proposed  using a formalism
borrowed from low-energy hadron physics and called Chiral Perturbation Theory
($\chi$PT) [7]. This approach has proved to be quite useful not only for the
model independent description of longitudinal electroweak gauge bosons
scattering (assuming the validity of the ET) [8] but also for the analysis of
the precision measurements  obtained at LEP [9].

To apply $\chi$PT to the description of the SBS of the SM one assumes that
there must be some  physical system coupled to the SM with a global symmetry
breaking from a group $G$ to another group $H$ which drives the spontaneous
symmetry breaking of the gauge group $SU(2)_L\times U(1)_Y$ to $U(1)_{em}$
thus giving masses to the the $W^{\pm}$ and $Z$ gauge bosons through the
standard Higgs mechanism. The  GB related to the global $G$ to $H$ symmetry
breaking live in the coset space $G/H$ and their low energy dynamics is
described by a gauged Non-Linear Sigma Model supplemented with  and infinite
number of higher derivative terms (but finite for practical applications)
needed for the renormalization of the model.

As we said above $\chi $PT has been used together with the ET to describe the
scattering of the longitudinal components of the gauge bosons. However, no
rigorous proof of this theorem has been presented until the present moment
outside the framework of the MSM. Moreover, it has been realized recently that
even in that case the simple original formulation of the SM should be
corrected in order to take into account the different renormalization of the
GB and the gauge bosons [10].

In this work we derive the general form of the theorem valid for a chiral
lagrangian description of the SBS of the SM including those renormalization
factors mentioned before. In so doing we complete the work presented in [11]
where we derived the theorem for regularized Green functions but not for
renormalized Green functions. Our derivation is based on the nice formal proof
of the ET for the MSM by Chanowitz and Gaillard [2] which relies in the
 BRS symmetry [12] of the Green functions, but we take into account
 the peculiarities  of $\chi $PT and include all the renormalization factors.
To implement the BRS symmetry we follow a very general methods given in
 [13] that provides a very useful way to build, starting
from a gauge invariant lagrangian, a quantum lagrangian  which is (anti)-BRS
invariant as well as a generalization of the standard Faddeev-Popov method,
and therefore, valid for the definition of $\chi $PT when the gauge fields are
quantized and included in loops. Note that this is not the case in the
original applications of $\chi$PT to the description of the low-energy pion
dynamics.

\section{The chiral lagrangian description of the SBS}
 In order to choose the
appropriate $G$ and $H$ groups we require the following conditions:  a) $dim
K=dim G/H=3$ as we need three GB to give mass to the three observed gauge
bosons $W^{\pm}$ and $Z$; b) $G$ should contain the $SU(2)_L\times U(1)_Y$
group so that the symmetry breaking sector can couple to the electroweak gauge
bosons; c) Since we want to ensure the experimental relation $\rho \simeq 1$,
we will require the subgroup $H$ to contain the custodial group $SU(2)_{L+R}$
[14]. This automatically yields $\rho =1$ when the gauge fields are switched
off and also implies that the photon will remain massless since $U(1)_{em}$ is
contained in $SU(2)_{L+R}$ and therefore in $H$. In [11] it was shown that
these conditions completely determine the $G$ and $H$ groups to be
$G=SU(2)_R\times SU(2)_L$ and $H=SU(2)_{L+R}$ so that $K=G/H=S^3$.

Therefore, the most general  dynamics of the symmetry breaking sector of the SM
compatible with  all known constrains can be defined as a gauged non-linear
sigma model based on the coset space $K=G/H=SU(2)_L\times
SU(2)_R/SU(2)_{L+R}=S^3$ with gauge group  $SU(2)_L\times U(1)_Y$. The
corresponding lagrangian can be written as:
\begin{eqnarray}
 {\cal L}_g    &=&{\cal L}_{YM}^L+{\cal L}_{YM}^Y  \\ \nonumber
&+&\frac{1}{2}g_{\alpha\beta}(\omega)D_{\mu}\omega^{\alpha}
D^{\mu}\omega^{\beta} \\ \nonumber
 &+& higher \; covariant \; derivative \; terms
\end{eqnarray}
where ${\cal L}_{YM}^L$ and ${\cal L}_{YM}^Y$ are the
Yang-Mills lagrangians for the
 $SU(2)_L$ and $U(1)_Y$ gauge fields $W_{\mu}^a$ and $B_{\mu}$,
$\omega^{\alpha}$
 are arbitrary coordinates chosen on the coset $S^3$ with coordinates
 $\omega^ \alpha=0$ for the classical vacuum. The action of the  group
 $G$ on this space defines the killing vectors $\xi^{\alpha}_{\;a}$
 through $\delta \omega^{\alpha}=\xi^{\alpha}_{\;a}(\omega)
\epsilon^a$ which is the non-linear transformation of the GB
$\omega^{\alpha}(x) $ under the action of an infinitesimal $G$ element.
Here, the $a$ index
runs from $1$ to $6$ where the values $1$ to $3$ correspond to
the unbroken $H=SU(2)_L\times SU(2)_R$ generators. The $S^3$
metrics $g_{\alpha\beta}$ is defined as follows: Let
  $e_a=e^{\alpha}_{\;a} \partial/\partial \omega^{\alpha}$
with $e^{\alpha}_{\;a}=\xi^{\alpha}_{\;a+3}$ for $a=1,2,3$ i.e. the dreibein is
 just the set of killing vectors corresponding to the $3$ broken
generators, then $g_{\alpha\beta}$ is the inverse of
$g^{\alpha\beta}$ where $g^{\alpha\beta}=e^{\alpha}_{\;a}e^{\beta a}$. It is
easy to show that $G$ is the isometry group of $S^3$ so that
$g'_{\alpha\beta}(\omega)=g_{\alpha\beta}(\omega)$ under any $G$
transformation.
The covariant derivatives are defined as:
\begin{equation}
D_{\mu}\omega^{\alpha}=\partial_{\mu}\omega^{\alpha}-gl^{\alpha}_{\;a}W_{\mu}^a-g'y^{\alpha}B_{\mu}
\end{equation}
where $l^{\alpha}_{\;a}$ and $y^{\alpha}$ are the killing vectors corresponding
to the gauged group $SU(2)_L\times U(1)_Y$. The {\it higher derivative terms}
include in principle
any covariant (in the space-time and the $S^3$ sense) piece with an arbitrary
high
number of covariant derivatives and arbitrary couplings so that
we can reproduce any dynamics compatible with the symmetry breaking pattern $
SU(2)_L\times SU(2)_R/SU(2)_{L+R}=S^3$ and the gauge group
$SU(2)_L\times U(1)_Y$. The gauge transformations are:
\begin{eqnarray}
\delta\omega^{\alpha}&=&l^{\alpha}_{\;a}\epsilon^a_L(x)+y^{\alpha}\epsilon_Y(x)
\\ \nonumber
 \delta W_{\mu}^a&=&
\frac{1}{g}\partial_{\mu}\epsilon^a_L(x)+\epsilon_{abc}
W_{\mu a}\epsilon_{Lc}(x) \\ \nonumber \delta
B_{\mu}&=&\frac{1}{g'}\partial_{\mu}\epsilon_Y(x)
\end{eqnarray}
\section{The BRS transformations and the quantum lagrangian}
The above gauge transformations
satisfy the well known properties of  closure
(since the Lie brackets of the gauged Killing vectors satisfy the
corresponding $SU(2)_L\times U(1)_Y$ algebra) and the Jacobi identity.
Following  [13] this fact makes it possible to build the related
(anti)-BRS transformations
by introducing the anti-commuting ghost fields $c_a$ and  $\bar c_a$,
and the commuting auxiliary field $B_a$  with $a=1,2,3,4$. In the
following we will use an unified notation where the first three values of
the gauge indices $a=1,2,3$ refer to the $SU(2)_L$ group and $a=4$ refers to
$U(1)_Y$ so that the gauge
field $W_{\mu}^a$ with $a=1,2,3,4$ will be defined as $W_{\mu}^a=W_{\mu}^a$
for
$a=1,2,3$ and $W_{\mu}^4=B_{\mu}$. In addition
we introduce the Killing vector $L^{\alpha}_{\;a}$
with $a=1,2,3,4$ as $L^{\alpha}_{\;a}=gl^{\alpha}_{\;a}$ for $a=1,2,3$ and
$L^{\alpha}_{\;4}=g'y^{\alpha}$ and the completely antisymmetric
symbols $f_{abc}$ with $a=1,2,3,4$ as
 $f_{abc}=g\epsilon_{abc}$ for $a=1,2,3$ and $f_{ab4}=0$.

 The
 closure relation and the Jacobi identity are equivalent
 to the nilpotency  properties of the $(\bar s)-s$ of the corresponding
(anti)-BRS transformations obtained using the general method of [13]
 \begin{equation}
 s^2=s \bar s+\bar s s=\bar s ^2=0
\end{equation}
These nilpotency properties are very important to define a quantum lagrangian
(anti)-BRS invariant as
\begin{equation}
{\cal L}_Q={\cal
L}_g+\frac{1}{2}s \bar s[W_{\mu}^aW^{\mu a}+2\xi f(\omega)+\xi c^a\bar c_a]
\end{equation}
where $f$ is any scalar analytical function with $\partial f(\omega)/\partial
\omega^{\alpha}=\omega^{\alpha}
+O(\omega^2)$. The new term added to the gauge invariant lagrangian
(see [11] for complete display) can be understood as a generalization of the
more usual
gauge fixing and Faddeev-Popov terms corresponding to the t'Hooft like gauges
($R_{\xi}$  gauges)
which have two main advantages: First, they provide a contribution to the
quadratic part in the gauge fields of the lagrangian so that the corresponding
operator can be inverted giving rise to well defined  $R_{\xi}$
propagators to be used in perturbation theory. Second, these gauges cancel the
unwanted  GB  and gauge boson mixing terms appearing in the
third term on eq.1  through the covariant derivatives. In addition, this
generalized method
 produces other GB-gauge boson and ghost-gauge boson
interactions. For gauges different from that of Landau ($\xi=0$) we also have
quartic ghost interactions and GB-ghosts interactions.

{}From the (anti)-BRS invariance of the lagrangian in eq.5 it is possible to
derive
the corresponding Ward identities for the dimensionally
regularized Green functions that are used in the proof of the ET. Note that the
use
of dimensional regularization is needed not only to preserve the (anti)-BRS
invariance in the regularized lagrangian but also to avoid the
$-\frac{i}{2}\delta^n(0)tr\; \log g$ term that would
otherwise appear in the quantum lagrangian of the non-linear sigma model (NLSM)
coming
from the path integral measure of the GB fields [15] .

However, in practice one is not only interested in the regularized
Green functions but also in the renormalized Green functions in order to make
predictions for the different physical processes. To cancel all the divergences
appearing in the Green functions obtained from the lagrangian in eq.5 one needs
to consider the renormalized lagrangian which consists on that of eq.5  plus
other terms with the corresponding couplings needed to reproduce all the
divergent structures appearing in the Green functions. The precise form of
these
counterterms is not known beyond those with four derivatives [16], but in any
case, they should be (anti)-BRS invariant too. Otherwise, the gauge invariance
of the model would be anomalous, i.e., broken by quantum effects. However, even
when we have chiral fermions coupled to GB and gauge bosons, it
is well known that the standard hypercharge assignments in the SM and the fact
that the number of colors is $N_c=3$ are such that
all possible gauge and  mixed gauge-gravitational anomalies cancel, including
the
non-perturbative
 $SU(2)$ discovered by Witten [17]. In addition there are potential
reparametrization invariance anomalies that could break the invariance under
changes of coordinates on the coset, but as it was shown in [18] these
anomalies
are absent from NLSM defined on spaces of dimension
lower than the space-time dimension as it happens in the case here considered.

	Therefore, when we take into account all the terms needed, we obtain  a
lagrangian with an infinite number of terms which is (anti)-BRS invariant. It
can be understood as the renormalized lagrangian of a renormalizable theory
(but
with an infinite number of couplings) written in terms of the bare fields an
couplings. We can also give this renormalized lagrangian using the renormalized
fields and couplings. The terms appearing in this case have the same form
(as the theory is renormalizable in the generalized sense described above) but
now some $Z$ factors appear in front of them. The relation between
the renormalized and the bare fields and gauge couplings is given by:

\begin{eqnarray}
W_{0\mu}^a(x) =Z_3^{(a)1/2}W_{\mu}^a(x) ;
\pi_o^{\alpha}(x)=Z_{\pi}^{(\alpha)1/2}\pi^{\alpha}(x) ;
g_0^{(a)}=Z_g^{(a)}g^{(a)} ;
\xi_0^{(a)} =Z_3^{(a)}\xi^{(a)}           \\   \nonumber
c_0^a(x) =\widetilde Z_2^{(a)1/2}c^a(x) ;
\bar c_0^a(x)=\widetilde Z_2^{(a)1/2} \bar c^a(x) ;
B_0^a(x)=\widetilde Z_2^{(a)}B^a(x)
; v_0 =Z_v^{1/2}v
\end{eqnarray}
where $g^{(a)}=g$ for $a=1,2,3$ and $g^{(4)}=g'$. The first three $Z_3$ are
equal due to the gauge structure of the model. Note that from now on we
 use as a
notation that the indices between parenthesis are not summed and that those
fields and constants without $0$ subscripts refer to the renormalized ones.
In addition we also
have infinite relations between the bare and the renormalized couplings
appearing in the chiral lagrangian. Due to the existence of an (anti)-BRS
symmetry in the renormalized lagrangian written in terms of the bare quantities
it is possible to find the corresponding "renormalized" (anti)-BRS
transformations which will  leave invariant the renormalized lagrangian once
written in terms of the renormalized fields and couplings. Those (anti)-BRS
transformations can be written as:
\begin{eqnarray}
s_R[\omega^{\alpha}]=X^{(a)}L^{\alpha}_{Ra}c^a \;\;\;\;\;\; &\qquad&
\bar s_R[\omega^{\alpha}]=X^{(a)}L^{\alpha}_{Ra}\bar c^a \\  \nonumber
s_R[W^{\mu a}]=X^{(a)}D^{\mu a}_{Rc}c^c \;\;\; &\qquad&
\bar s_R[W^{\mu a}]=X^{(a)}D^{\mu a}_{Rc}\bar c^c \\  \nonumber
s_R[c^a]=-\frac{X^{(a)}}{2}f^a_{R\;bc}c^bc^c &\qquad&
\bar s_R[c^a]=-\frac{X^{(a)}}{2}f^a_{R\;bc}\bar c^bc^c \\ \nonumber
s_R[\bar c^a]=X^{(a)}\frac{B^{(a)}}{\sqrt{\xi^{(a)}}} \;\;\;\;\;\;\; &\qquad&
\bar s_R[\bar c^a]=-X^{(a)} \left( \frac{B^{(a)}}{\sqrt{\xi^{(a)}}} +
f^a_{R\;bc}\bar c^bc^c \right) \\ \nonumber
s_R[ B^a]=0 \;\;\;\;\;\;\;\;\;\;\;\;\;\;\;\;\;\;\; &\qquad&
\bar s_R[ B^a]=-X^{(a)} f^a_{R\;bc}\bar c^bB^c \\  \nonumber
\end{eqnarray}

 where $L^{\alpha}_{Ra}=Z_{\pi}^{(\alpha)-1/2}Z_3^{(a)1/2}L^{\alpha}_{\;a}$
 and $f^a_{R\;bc}=Z^{(a)}_g Z_3^{(a)1/2} g f^a_{\;bc}$.

 Once we have a set of (anti)-BRS
 symmetry transformations for the renormalized lagrangian
 in terms of the renormalized fields we can apply standard  methods
 to obtain the corresponding Ward identities
 for the renormalized Green functions. In particular, some of those
relations will be used to find the version of the ET that
 applies to the chiral lagrangian description of the symmetry breaking
 sector of the SM.

\section{Ward Identities}

	The ET provides the relationship between $S$-matrix elements involving an
arbitrary number of longitudinal gauge bosons $W_L$ and those elements with
 all the external $W_L$ replaced by their corresponding GB. To find that
 relation, and
following the steps of the Chanowitz-Gaillard proof [2], we will make use of
the renormalized lagrangian BRS invariance to derive Ward identities relating
the desired renormalized connected Green's functions, that will be converted
 in relations
between $S$-matrix elements by means of the
 Lehmann-Symanzik-Zimmermann (LSZ) reduction formula.

	Indeed, the renormalized effective action $\Gamma_R[A]$ obtained from the
quantum lagrangian of the gauged NLSM is (anti-) BRS invariant too.
($A$ stands for any field appearing in the quantum lagrangian i.e.
$A_i=\omega^{\alpha}, W^a_{\mu}, c^a, \bar c^a, B^a$). This invariance can be
stated as follows:
\begin{equation}
\sum_i \int d^4x s_R[A_i]\frac{\delta \Gamma_R[A]}{\delta A_i}=0
\end{equation}
	The generating functional for renormalized connected Green's functions
\newline
 $W_R(x_1,...,x_n)$  is given, in momentum space, by the following definition:
\begin{equation}
W_R[J]=(2\pi)^4 \sum_{n=1} \int \prod_{i=1}^n \frac{d^4p_i}{(2\pi)^4}
\delta^4(\sum_i p_i)J_{i_1}(-p_1)...J_{i_n}(-p_n) W_{
R \; i_1,...,i_n}(p_1,...p_n)
\end{equation}
Using the well known relations:
\begin{equation}
 W_R[J]+\Gamma_R[A]+\sum_i \int d^4x J_i(x) A_i(x)=0; A_i(x)=-\frac{\delta
W_R[J]}{\delta J_i(x)}; J_i(x)=-\frac{\delta \Gamma_R[A]}{\delta A_i(x)}
\end{equation}
we can write the fields as follows:
\begin{equation}
 A_i(k)=(2\pi)^4 \sum_{n=1} \int \prod_{j=1}^n
 \frac{d^4p_j}{(2\pi)^4} J_{i_j}(-p_j) W_{R \; i,i_1,...,i_n}(k,p_1,...p_n)
\delta^4(k+\sum_j p_j)
\end{equation}
It is important to notice that all the terms in this expansion contain at
least one current $J$. Gathering these expressions, the condition of BRS
invariance for the effective action is written:
\begin{equation}
 \sum_i \int \frac{d^4k}{(2\pi)^4} s_R[A_i(k)] J_i(-k)=0
\end{equation}
	It is now straightforward to obtain Ward identities just by expanding the
fields contained in $s_R[A_i(k)]$ in terms of connected Green's functions as
in eq.11, and then taking functional derivatives with respect to $J_i(p)$ at
$J=0$.  Since we are mainly interested in relations concerning $W_L$ and GB, we
need to consider Green functions involving the auxiliary $B$ field, as it is
nothing but the gauge fixing condition that intuitively identifies $W_L$ and
the GB.
	To illustrate the general procedure we derive now a Ward identity for
the two-point Green function with one $B$ field. The result thus obtained
will be used later in the complete proof for the modified ET. So we write:
\begin{equation}
 \left.\frac{\delta}{\delta
J_{\bar c_b}(-k)}\frac{\delta}{\delta J_j(p)} \sum_i \int \frac{d^4k}{(2\pi)^4}
s_R[A_i(k)] J_i(-k)\right|_{J=0} =0
\end{equation}
 The only possible contributions come from those terms in the BRS
transformations involving just one current. Thus, from eqs.7  and 11 we obtain:
\begin{equation}
 \frac{X^{(b)}}{\sqrt{\xi^{(b)}}}W_{B^{b}l}(p)
=- X ^{(a)}D^{a}_{ R l}(p) W_{c^{a}\bar c^{b}}(p) =
 \end{equation}
where:
\begin{equation}
D^{a}_{ R l}(p) = ip_{\mu}\delta_l^{W_{\mu a}}+L^{(0) \alpha}_{R a}
\delta_l^{\omega^{\alpha}}
 \end{equation}
	Here we can see two important differences with the formal proof of [2]: First,
the $X$ factors coming from renormalization. Second, the $L_R$ term which is
due to the nonlinear realization of the symmetry. Nevertheless, this term only
contributes in the zeroth order $L_R^{(0)}$, thus simplifying the complicated
relation between gauge  and Goldstone bosons that one would expect naively from
the nonlinear gauge fixing condition.
	We now want to obtain the general expression, and so we start from:
\begin{equation}
 \left.\frac{\delta}{\delta J_{\bar c_{a_1}}(-k)}
\prod_{j=2}^{s}\frac{\delta}{\delta J_{B_{a_j}}(-k_j)}
\prod_{k=1}^{m}\frac{\delta}{\delta J_{A_k}(-p_k)} \sum_i \int
\frac{d^4q}{(2\pi)^4} s_R[A_i(q)] J_i(-q)\right|_{J=0} =0
\end{equation}
 Where we will impose that the currents $J_{A_k}$ are only
associated to physical $A_k$ fields. We can easily see from the BRS
transformations that we do not get any contribution if $A_i=B$,
neither when $A_i=\omega,c$ since there are no $J_{\omega}$ nor $J_c$
derivatives. Since the $A_k$ are physical, their polarization vectors will
cancel the derivative term in $s_R[W^a_{\mu}] = ik_\mu
c^{a}+\epsilon^a_{Rbc}W_{\mu b}c_c$ because $\epsilon \cdot k_\mu =0$. Thus, we
are only considering those terms coming from $s_R[\bar c]$ and the part which
is
left from $s_R [W^a_{\mu}]$ that we will call, generically, "bilinear terms".
Finally we obtain:
\begin{equation}
 \frac{X^{(a_1)}}{\sqrt {\xi^{(a_1)}}}
W_{B_{a_1}B_{a_2}...B_{a_s}A_1...A_m}(k_1,...,k_s,p_1,...p_m) + \mbox{
bilinear terms} =0
\end{equation}
where $\sum_i k_i =-\sum_i p_i$. In order to translate this result to off-shell
$S$-matrix elements, we apply the LSZ reduction formula:
\begin{eqnarray}
  \frac{X^{(a_1)}}{\sqrt {\xi^{(a_1)}}} \left( \prod_{i=1}^{m}W_{A_iA_i}
(p_i) \right) \sum_{l_j} \left( \prod_{j=1}^{s}W_{B_{a_j}l_j}(k_j) \right)
S^{off-shell}_{l_1..l_s A_1...A_m}(k_1...k_s,p_1...p_s)  \\ \nonumber
 + \mbox{ bilinear terms}=0
\end{eqnarray}

	As the $a_1$ index is free we can drop the factor $X/ \sqrt{\xi}$ which is
irrelevant. When we
multiply the whole last equation by the inverse $A_i$ propagators,
setting their momenta on shell, that is $p^2_i=m_{A_i}^2$, we can use  the
same argument as in [2] to cancel the "bilinear terms" since they are the
product of two connected Green's functions contracted properly, but with
one off-shell momentum, and without the pole needed to compensate for
$W_{A_iA_i}^{-1}(p_1) \rightarrow 0$ when $p_1^2=m_{A_1}^2$. Therefore,
using eq.14 to substitute the $B$ field two point functions, we obtain:

\begin{equation}
\left. \sum_{l_j} \left(\prod_{j=1}^{s}\frac{\sqrt {\xi^{(a_j)}}}{X^{(a_j)}}
X^{(c_j)} W_{c^{c_j}\bar c^{a_j}}(k_j) D^{c_j}_{ R l_j}(k_j) \right)
S^{off-shell}_{l_1..l_s A_1...A_m}(k_1...k_s,p_1...p_s)
\right|_{p^2_i=m_{A_i}^2} =0
\end{equation}
Now, we can take away the $\sqrt {\xi^{(a_j)}}/X^{a_j}$ factors since the $a_j$
 are not contracted.
Then, we multiply by the ghost (non-diagonal, in principle) inverse two point
functions $W^{-1}_{c^{d_j}\bar c^{a_j}}(k_j)$ so that the $d_j$
index is again free allowing us to drop the last $X$ factor. Therefore, we
arrive to the following expression:
\begin{equation}
\left. \sum_{l_1...l_r}\prod_{i=1}^{s}D^{a_i}_{R l_i}(p_i) S^{off-shell}
_{l_1..l_s,A_1..A_m}(p_1..p_r,k_1..k_m) \right|_{p^2_i=m_{A_i}^2} =0
\end{equation}

\section{The Equivalence Theorem}

Our aim is to obtain the $S$-matrix elements from this formulae by
setting all the momenta on-shell, but the $W_{\mu}$ fields in the $D_R$
operator are not physical fields. We still have to obtain the
physical combinations by means of a transformation  $\widetilde
W^a_{\mu}= R^{ab}W^b_{\mu}$, whose most general form will be :
\begin{equation}
\left ( {\matrix{ \widetilde W_{\mu}^1 \cr
\widetilde W_{\mu}^2 \cr
\widetilde W_{\mu}^3 \cr
\widetilde W_{\mu}^4  }} \right ) =
\left ( {\matrix{ W_{\mu}^- \cr W_{\mu}^+ \cr Z^{phys}_{\mu} \cr A^{phys}_{\mu}
}}
\right) = \left ( {\matrix{ 1/\sqrt{2}&i/\sqrt{2}&0&0\cr
1/\sqrt{2}&-i/\sqrt{2}&0&0\cr
0&0&cos \theta & -sin \theta \cr
0&0&sin \theta' & cos \theta'  }} \right )
\left ( {\matrix{ W_{\mu}^1 \cr W_{\mu}^2 \cr W_{\mu}^3 \cr W_{\mu}^4
 }} \right)
\end{equation}
These new fields are the renormalized fields which ensure that the poles of the
exact propagators are located at the values of the corresponding physical
masses
 . Once we have obtained them, we also
define: $\widetilde L^{(0)a}_{R\alpha}=L^{(0)a}_{R\alpha}
(R^{-1})^{ba}$. Therefore we finally write:
\begin{equation}
\sum_{l_1...l_r}\prod_{i=1}^{s}\widetilde{D}^{a_i}_{R l_i}(p_i)
S _{l_1..l_s,A_1..A_m}(p_1..p_r,k_1..k_m) = 0
\end{equation}
where
\begin{equation}
\widetilde{D}^{a}_{ R l}(p) = ip_{\mu}\delta_l^{\widetilde{W}_{R \mu
a}}+\widetilde{L}^{(0) \alpha}_{ Ra} \delta_l^{\omega^{\alpha}}
\end{equation}
and we have set the $p_i$ momenta on-shell for the massive physical vector
bosons.

 The next step to obtain the modified version of the ET is
to substitute the momenta in each $\widetilde{D}^{a}_{Rl}(p)$ using the
relation $\epsilon_{(L)\mu}=p_\mu / m+v_\mu$ and then neglect at high energies
those terms containing $v_\mu$ factors since $v_\mu \simeq
O(M_{phys}/E)$. Unfortunately, this is not a straightforward procedure due to
the gauge structure of the theory which is responsible for cancellations  in
those amplitudes involving longitudinally polarized gauge bosons, and does not
allow us to simply neglect the terms containing $v_{\mu}$ factors.

	We can go around this problem using the following relation between amplitudes
that we will write symbolically as:
\begin{eqnarray}
\left( \prod_{i=1}^{n}\epsilon_{(L)\mu_i}\right)
T(\widetilde W^{\mu_1}_{a_1},..., \widetilde W^{\mu_n}_{a_n};A) = \hspace {8cm}
 \\ \nonumber
 = \sum_{l=0}^{n} (-i)^l \left( \prod_{i=1}^{l}v_{\mu_i} \right) \left(
\prod_{j=l+1}^{n}K^{a_j}_{\alpha_j} \right) \bar T(\widetilde W^{\mu_1}_{a_1}
...\widetilde W^{\mu_l}_{a_l},\omega_{\alpha_{l+1}} ...\omega_{\alpha_n};A)
\end{eqnarray}

  where we have omitted the irrelevant indices which are supposed to be
properly contracted, and we have defined $K^{a}_{\alpha}=\widetilde{L}^{(0)
\alpha}_{Ra} / M^{(a)}_{phys}$. In the right hand side we write $\bar T$ since
for each $l$ value we  carry a sum over all the amplitudes with independent
permutations of fields and indices. A very similar relation, but without
considering the $K$ factors, was first obtained in [2], the derivation of the
formula taking them into account is completely analogous and we do not
reproduce it here. These $K$ will modify the  final statement of the ET.

	When the amplitudes satisfy the unitarity bounds, we can drop
at high energies all terms in the RHS of eq.24 but the one with $l=0$
which is precisely that with all external $\widetilde W_L$ substituted by
GB. This step is allowed since the amplitudes will never grow with the energy
and therefore those terms which contain $v_{\mu}$ will vanish in the high
energy
limit. However, when considering effective lagrangians, the amplitudes are
obtained
perturbatively as a truncated series in the energy so that the same reasoning
is no
longer valid, and we have to use power counting methods to extract the leading
contributions.

 In this case  we can, in principle, expand the
amplitudes as Laurent series in $E/4\pi v$ up to a positive power $N$ by
fixing the maximum number of derivatives in the Lagrangian. However, as we will
require these
amplitudes to satisfy the Low Energy Theorems (second reference in [14]) in the
$M^2 \ll E^2$ regime, we can write the energy negative powers as $(M/E)^{k}$
 (To simplify the analysis we have set momentarily $g'=0$). Thus we write:
\begin{eqnarray}
 \bar T(\widetilde W^{\mu_1}_{a_1}
...\widetilde W^{\mu_l}_{a_l},\omega_{\alpha_{l+1}} ...\omega_{\alpha_n};A)
&\simeq&
\sum_{k=0}^{N} a_{l}^k \left(\frac{E}{4\pi v} \right) ^k + \sum_{k=1}^{\infty}
a_{l}^{-k} \left( \frac{M} {E}\right)^{k} \\ \nonumber
 ( \prod_{i=1}^{n}\epsilon_{(L)\mu_i} )
 T(\widetilde W^{\mu_1}_{a_1},..., \widetilde W^{\mu_n}_{a_n};A)
&\simeq& \sum_{k=0}^{N}
b^k  \left( \frac{E}{4\pi v} \right)^k + \sum_{k=1}^{\infty} b^{-k}
\left( \frac{M}{E} \right)^{k}
\end{eqnarray}
	These series are formal since the  coefficients can contain energy logarithms
(for the sake of brevity we have omitted the field indices in $a^k_l$ and
$b^k$).
Furthermore, these coefficients can be expanded perturbatively on
$g$, for instance: $a_{l}^h=a_{lL}^h(1+ O(g/4\pi))$ where $a_{lL}^h$ is the
lowest
order term in the expansion of $a^h_l$ in powers of $g$.
 In most renormalization schemes we have $M \simeq
M_{phys} (1+O(g/4\pi) )$ which means that we can write as well
$K^a_{\alpha}\simeq K^{a(0)}_{\alpha}+ K^{a(1)}_{\alpha} (g/4\pi)+....$ where
these coefficients are energy independent. Introducing these expansions in
eq.24 and neglecting terms of order $O(M/E)$ and $O(E/4\pi v)^{N-n+1}$, we
arrive to the following expression:
\begin{eqnarray}
 ( \prod_{i=1}^{n}\epsilon_{(L)\mu_i} )
T(\widetilde W^{\mu_1}_{a_1},..., \widetilde W^{\mu_n}_{a_n};A)  \simeq
\hspace{8cm} \\ \nonumber
\simeq \left(\prod_{j=1}^{n}K^{a_j (0)}_{\alpha_j} \right)
\sum_{k=0}^{N-n} (a_{0L}^{k}(1 + O(g/4\pi )))  \left( \frac{E}{4\pi v}
\right)^k
+O\left( \frac{M}{E} \right)+O\left( \frac{E}{4\pi v} \right)^{N-n+1}
\end{eqnarray}
 which is the statement of the ET for chiral lagrangians (Note that we have
omitted for brevity the indices $\alpha$ of $a_0$). In fact, if we want these
approximations to make sense, we have to restrict the values of the energy to
the
 following applicability window:
\begin{eqnarray}
 M \ll E \ll 4\pi v= 4 \pi M/g \\ \nonumber
 g/4\pi \ll (E/4\pi v)^{N-n+1}
\end{eqnarray}
The first two inequalities come from neglecting the $O(M/E)$ and
$O(E/4\pi v)^{N-n+1}$ terms respectively. The last constraint is needed since
we are
 taking into account the $O(E/4\pi v)^{N-n}$ contribution while neglecting that
of
$O(M/E)$, therefore we expect the former to be much bigger than the latter.

The generalization to the $g' \neq 0$ is straightforward due to the fact that
$g'
\ll g$ as well as $ M_Z ^{phys} \simeq M_W ^{phys} \simeq M^{(a)}$ for any $a$
( all the different masses are of the same order when counting energy powers):
we only consider the lowest order of the $a$ coefficients in the $g$ or $g'$
expansion so that
the same reasoning we had used when $g' = 0$  is still valid. Thus we can use
eq.26 as the precise formulation of the ET for chiral lagrangians.

	It is important to remark that this version of the ET states  that if we
expand in terms of the energy the $S$ matrix element of a process
 involving $W_L$ bosons, and we do the same with the $S$ matrix
element obtained by substituting all the longitudinal gauge bosons by their
respective GB, we are allowed to approximate the coefficients of the first
series not by the corresponding coefficients of the second one, but by
their lowest order in terms of $g$ or $g'$. Even more, this approximation is
expected to be useful when the energy regime satisfies the constraints of
eq.27.

\section{Discussion}

In the preceding section we have stated a generalized version of the
ET which holds for the chiral lagrangian formalism. As a
matter of fact, eq.26 is valid, even when $g' \neq 0$,  for renormalized
amplitudes calculated at any
 order in the chiral expansion, any choice of the renormalized $\xi^{(a)}$
parameters and  any parametrization of the GB coset space.  The main
difference with the formal
 version of the ET for regularized amplitudes [11] is contained in the $K$
factors, one per GB, that multiply the RHS of eq.26, that are basically made
of the zeroth order of the killing  vectors, which selects the GB combination
to be eaten by the physical gauge fields,  and renormalization constants.

The demonstration has two separated parts: First, from the BRS invariance of
the renormalized lagrangian we derive Ward identities to relate renormalized
Green functions involving external longitudinal components of gauge  bosons to
those Green functions where we replace some of this external legs by their
corresponding GB as in eqs.14 and 17. The possibility to build a (anti-)BRS
invariant renormalized  lagrangian although we do not know the precise form of
the needed counterterms  is due to the absence of anomalies when quantizing
the theory. Even though we  considered chiral fermions, the usual hypercharge
assignments and the fact that $N_c=3$ cancel all possible
 gauge and gauge-gravitational anomalies, as well as the non-perturbative
$SU(2)$ discovered by Witten [17]. Some issues related with this point have
been recently been discussed in [19]. We have also given a formalism  which is
invariant under changes of coordinates in the GB coset, since not only have we
built a lagrangian invariant under such changes, but the possible
reparametrization anomalies are absent since the coset dimension is lower than
that of  the space-time [18]. Once we have these relations between Green
functions we translate them to relations among $S$ matrix elements through the
LSZ reduction formula as in eq.22, it is
 then necessary to rotate the renormalized fields in the lagrangian in order
to obtain  the physical combinations of fields and set the momenta on shell
for the gauge
 bosons.

Second, the identities thus obtained involve momentum factors that have to be
converted in polarization vectors using $\epsilon_{(L)\mu}= p_{\mu}/M+v_{\mu}$
by neglecting at high energies the  $v_{\mu}$ factor since it is $O(M/E)$.
This step is not straightforward  when we are dealing with chiral lagrangians,
since the amplitudes  are obtained as a chiral expansion in  $E/4\pi v$ and
they can grow with positive powers of the energy  violating the unitarity
bounds. Therefore  we have to perform a power counting analysis to drop the
$v_{\mu}$ terms, and obtain the generalized version of the ET. In order to
make compatible all the approximations, we have to restrict the possible
values of the energy. There will be a high energy bound needed to neglect the
$v_{\mu}$ factors as well as a low energy bound in order to make the chiral
expansion in energy powers.

 The apparent contradiction between these requirements could be
avoided thanks to the smallness of the couplings $g/4\pi$ and $g'/4\pi$ which
are involved in the expansion at low energies. The actual existence of an
applicability window for the theorem given a perturbative order for the
calculations can only be tested by detailed computations of different
scattering processes. Nevertheless, the validity range of the theorem could be
enlarged if one uses non-perturbative techniques to implement unitarity, which
would then allow us to avoid the power counting methods, and thus to eliminate
the high energy requirements. Among others, these techniques include:
dispersion relations and Pad\'e approximants [20],  and the large $N$ limit
[21], which have been found to work very well when applied to hadron physics.

\section{Conclusion}

	In this work we have stated the version of the ET valid for
the  $\chi$PT methods used to describe the GB and gauge
 boson dynamics of the symmetry breaking sector of the SM. Starting from a
gauged NLSM based on the coset $S^3=SU(2)_L \times SU(2)_R/SU(2)_{L+R}$, we
have built a renormalized quantum lagrangian which is not only (anti-)BRS
invariant  but also under reparametrizations of the coset. The existence of
these symmetries is due to the absence of anomalies in the theory, and they
provide us with Ward identities relating renormalized Green functions with
external longitudinally polarized gauge bosons and those with GB.

	Through a power counting analysis we have been able to formulate the ET
version that holds for the renormalized physical amplitudes when calculated by
means of a chiral expansion, for any coordinate choice of the GB and for any
values of the renormalized $\xi^{(a)}$ parameters. This version  states  that
when expanding in terms of the energy, we can approximate the  coefficients of
the $S$ matrix element of a process
 involving $W_L$ bosons, not by the corresponding coefficients of the
 $S$ matrix element obtained by substituting all the longitudinal gauge bosons,
but by
their lowest order in terms of $g$ or $g'$.
 This fact, and the appearance of the $K$ factors due to the renormalization
procedure together with the existence of an upper bound
in the energy applicability range, are the main differences with the original
formulation of the ET.

 In principle, the energy window given in eq.27 for applying simultaneously
$\chi$PT and the ET at a given loop level can be narrow  but the use of
non-perturbative techniques could extend it to higher energy regions thus
providing a model independent description of the SBS dynamics. Work is in
 progress in that direction.

\section{Acknowledgements}

 This work has been supported in part by the Ministerio de Educaci\'on y
Ciencia (Spain)(CICYT AEN90-0034). A.D. also thanks the Gregorio del Amo
foundation (Universidad Complutense de Madrid)  for support and S. Dimopoulos
and  the Department of Physics of the Stanford University for their kind
hospitality during the first part of this work.

\section{Note Added}

When this work was being completed we noticed the appearance of two
 related preprints on the subject. In [22] the authors arrive, up to the level
 of the identities obtained from the BRS invariance, to similar results
 to us for $g'=0$ and a particular parametrization of the GB. However, that
work has been critizised in [23] since their analysis does not
include the power counting subtleties that
arise when the amplitudes are given as truncated series in the energy, as it
is customary in the applications of $\chi$PT. Those
considerations
(see also H.Veltman in [10]) give rise to the upper bound in the
applicability range of the ET for chiral lagrangians that has been given in
the present work.



\thebibliography{references}

\bibitem {1}J.M. Cornwall, D.N. Levin and G. Tiktopoulos, {\em Phys.
Rev.}  {\bf D10} (1974) 1145    \\
	C.E.Vayonakis,{\em Lett.Nuovo.Cim.}{\bf 17}(1976)383 \\
 B.W. Lee, C. Quigg and H. Thacker, {\em Phys. Rev.} {\bf D16} (1977)
 1519

\bibitem{2} M.S. Chanowitz and M.K. Gaillard, {\em Nucl. Phys.} {\bf
B261}(1985) 379


\bibitem{3} G.J. Gounaris, R. Kogerler and H. Neufeld, {\em Phys. Rev.} {\bf
D34} (1986) 3257



\bibitem{4} H. Haber and G. Kane, {\em Phys.Rep.}{\bf 117} (1985) 75


\bibitem{5} E. Farhi and L. Susskind, {\em Phys. Rep.} {\bf 74 } (1981) 277 \\
     S. Dimopoulos and L. Susskind, {\em Nucl. Phys.} {\bf B155} (1979) 237


\bibitem{6} R.Casalbuoni,S.de Curtis,D.Dominici and R.Gatto, {\em Phys. Lett.}
{\bf B155}(1985)95;{\em Nucl. Phys.} {\bf B282} (1987) 235

\bibitem{7}  S. Weinberg, {\em Physica} {\bf 96A} (1979) 327 \\
  J. Gasser and H. Leutwyler, {\em Ann. of Phys.} {\bf 158}
 (1984) 142, {\em Nucl. Phys.} {\bf
B250} (1985) 465 and 517

 \bibitem{8}  A. Dobado and M.J. Herrero, {\em Phys. Lett.} {\bf B228}
 (1989) 495 and {\bf B233} (1989) 505 \\
 J. Donoghue and C. Ramirez, {\em Phys. Lett.} {\bf B234} (1990)361  \\
A. Dobado, M.J. Herrero and J. Terr\'on, {\em Z. Phys.} {\bf C50} (1991) 205
and  {\em Z. Phys.} {\bf C50} (1991) 465 \\
S. Dawson and G. Valencia, {\em Nucl. Phys.} {\bf B352} (1991)27

\bibitem{9}  B.Holdom and J. Terning,
{\em Phys.Lett.}   {\bf B247} (1990) 88\\
A. Dobado, D. Espriu and M.J. Herrero        {\em  Phys.Lett.}
{\bf B255} (1991) 405\\
M. Golden and L. Randall, {\em Nucl. Phys.} {\bf
B361} (1991) 3

\bibitem{10} Y.P.Yao and C.P. Yuan, {\em Phys. Rev.} {\bf
D38} (1988) 2237   \\
  J. Bagger and C.Schmidt, {\em Phys. Rev.} {\bf
D41} (1990) 264  \\
H. Veltman, {\em Phys. Rev.} {\bf
D41} (1990) 2294 \\
H.J. He, Y.P. Kuang and X. Li, {\em Phys. Rev. Lett.} {\bf
69} (1992) 2619  \\
W. B. Kilgore, {\em Phys.Lett.} {\bf
B294} (1992) 257

\bibitem{11} A. Dobado and J.R. Pel\'aez, Stanford  Preprint SU-ITP-93-33
, hep-ph 9401202

\bibitem{12} C. Becchi, A. Rouet and R. Stora, {\em Comm. Math. Phys.} {\bf
42}(1975)
127

\bibitem{13} L. Baulieu and J.Thierry-Mieg {\em Nucl. Phys.} {\bf B197} (1982)
 477 \\
L. Alvarez-Gaum\'e and L. Baulieu, {\em Nucl. Phys.} {\bf B212} (1985) 255 \\
L. Baulieu, {\em Phys. Rep.} {\bf 129 } (1985) 1


\bibitem{14} P. Sikivie et al., {\em Nucl. Phys.} {\bf
B173} (1980) 189\\
M. S. Chanowitz, M.  Golden and H. Georgi
	      { \em Phys.Rev.}  {\bf D36}  (1987)1490


\bibitem{15} J. Charap, {\em Phys. Rev.} {\bf D2} (1970)1115  \\
 I.S. Gerstein, R. Jackiw, B. W. Lee and S. Weinberg, {\em Phys. Rev.}  {\bf
D3} (2486)1971\\
J. Honerkamp, {\em Nucl. Phys.} {\bf B36} (1972)130  \\
{\it Quantum Field Theory and Critical Phenomena},
  J. Zinn-Justin, Oxford University Press, New York, (1989)  \\
  L. Tararu, {\em Phys. Rev.} {\bf D12} (1975)3351  \\
 D. Espriu and J. Matias, Preprint UB-ECM-PF 93/15


\bibitem{16} T. Appelquist and C. Bernard, {\em Phys. Rev.} {\bf D22} (1980)
 200 \\
A. C. Longhitano, {\em Nucl.Phys.} {\bf B188} (1981)   118


\bibitem{17} E. Witten,{\em Phys. Lett.} {\bf B117} (1982)324


\bibitem{18}L. Alvarez-Gaum\'e and P. Ginsparg,
	    {\em Nucl.Phys.} {\bf B262}(1985) 439

\bibitem{19} J.F.Donoghue,{\em Phys. Lett.} {\bf B301} (1993)372 \\
P.B.Pal,{\em Phys. Lett.} {\bf B321} (1994)229 \\
W.B.Kilgore, {\em Phys. Lett.} {\bf B323} (1994)161



\bibitem{20} Tran N. Truong, {\em Phys. Rev.} {\bf D61} (1988)2526\\
  A. Dobado, M.J. Herrero and J.N. Truong, {\em Phys.
 Lett.} {\bf B235}  (1990) 134  \\
T.N.Truong, {\em Phys. Rev. Lett.} {\bf 67} (1991)2260  \\
 A. Dobado and J.R. Pel\'aez, {\em Phys. Rev.} {\bf D47}(1992)4883

\bibitem{21} C.J.C. Im, {\em Phys. Lett.} {\bf B281} (1992)357\\
 A. Dobado and J.R. Pel\'aez, {\em Phys. Lett.} {\bf B286}  (1992)136\\
 M. J. Dugan and M. Golden,{\em Phys.Rev.} {\bf D48} (1993)4375

\bibitem{22}H.J.He, Y.P.Kuang, and X.Li, Tsinghua preprint TUIMP-TH-94/56,
hep-ph/9403283

\bibitem{23} C.Grosse-Knetter, I.Kuss. Bielefield preprint BI-TP 94/10,
hep-ph/9403291
\end{document}